\documentstyle[aps,multicol,psfig]{revtex}
\begin{document}
\preprint{SNUTP }
\draft
\title{Pinning/depinning of crack fronts in heterogeneous materials \\}

\author{P.~Daguier$^{\dag}$, B.~Nghiem$^{\ddag}$, E.~Bouchaud$^{\dag}$ and
 F.~Creuzet$^{\ddag}$}
\address{
     \dag O.N.E.R.A. (OM), 29 Avenue de la Division Leclerc, \\
          B.P. 72, 92322 Ch\^atillon Cedex, FRANCE\\
     \ddag  Laboratoire CNRS/Saint-Gobain {\sl ``Surface du Verre
 et Interfaces"}, 39, Quai Lucien Lefranc,\\
         B.P. 135, 93303 Aubervilliers Cedex, FRANCE \\}

\maketitle

\begin{abstract}
The fatigue fracture surfaces of a metallic alloy, and the stress corrosion
fracture surfaces of glass are investigated as a function of crack velocity. 
It is shown that in both cases,
there are two fracture regimes, which have a well defined self-affine signature. 
At high enough length scales, 
the universal roughness index $\zeta \simeq 0.78$ is recovered. At smaller
length scales, the roughness exponent is close to $\zeta _c \simeq 0.50$.
The crossover length $\xi_c$ separating these two regimes strongly depends
 on the material, and exhibits a power-law decrease with the measured crack velocity $\xi_c \propto v^{-\phi }$, with $\phi \simeq 1$. The exponents $\nu$ and $\beta $ characterising the dependence of $\xi_c$ and $v$ upon the pulling force are shown to be close to $\nu \simeq 2$ and $\beta \simeq 2$.\\  
\end{abstract}


\begin{multicols}{2}
%\narrowtext

The pinning/depinning transition \cite{Fisher,Natt} has been the subject of many theoretical studies in the recent years \cite{PhysRep}. But although there are in principle numerous 
applications, experimental examples remain scarce. In this letter, we analyse experimental results concerning 
crack propagation in two very different materials $-$ a metallic alloy and glass $-$ within the framework 
of models of pinning/depinning of moving lines through randomly distributed obstacles \cite{ertaskardar1,ertaskardar3}.\par
Fracture surfaces of many heterogeneous materials have been studied, with different experimental 
techniques. As shown first by Mandelbrot {\sl et al} \cite{natureBM}, these surfaces are self-affine \cite{Mand}, 
with a roughness index $\zeta$ in most cases close to the value 0.8. This quantity was later conjectured to be universal \cite {alus,maloy}, i.e. independant of the material and of the fracture mode. As far as metallic materials are concerned, this exponent has been recently characterized over five decades of length scales (0.5 nm-0.5 mm) \cite {Daguier,MRS95}. This universality was first questioned by Milman {\sl et al} \cite {mil} on the basis of Scanning Tunneling Microscopy experiments where fracture surfaces of metallic materials were investigated at the nanometer scale. The reported values of roughness exponents in the latter case were significantly smaller than 0.8, closer to 0.5. This has been interpreted \cite {Daguier,EBSN} as a kinetic effect similar 
to the one expected for a moving line near its depinning transition. Indeed, it was proposed recently that a
fracture surface could be modelled as the {\it trace} left by the crack front moving through randomly distributed 
microstructural obstacles \cite {PRL}.\par

Various models have been developped to calculate the critical exponents characterising the morphology of these lines either in 2 dimensions (moving front) or in  3 dimensions (moving front leaving a surface behind it). Erta\c s and Kardar studied a local nonlinear three-dimensional Langevin equation
to describe the morphology of polymers in shear flows or the motion of flux lines in superconductors \cite {ertaskardar1}, and it was conjectured that these models might also be relevant for fracture\cite{PRL}. The line is pulled away with a constant force $F$. Non linearities account for the variations of the local crack
speed with the local orientation of the front. This equation leads to a large
 number of regimes,
depending on the relative values of the prefactors of
the non linear terms.
For some values, this model predicts that for a finite velocity $v$, the roughness exponent is 0.75 at
``large length scales" and 0.5 at ``short length scales", the two regimes being separated by a crossover length $\xi _c$. The short length scales regime corresponds to the vicinity of the depinning transition \cite{Fisher,Natt} where the crack front is just able to free itself from the pinning microstructural obstacles. In this case, i.e. when $F$ is close although higher than a critical 
force $F_c$ under which the line remains still, the velocity $v$ tends to zero, $v\propto (F-F_c)^{\beta }$, and $\xi _c$ diverges as $\xi _c \propto v^{-\phi}$. In this particular model \cite{ertaskardar1}, $\phi=3$. It will be shown in the following that the observed behavior is in qualitative agreement with this scenario \cite{Daguier}, although the measured value of $\phi$ is significantly smaller.\par
Note that this behavior also corresponds to the results of recent large-scale molecular dynamics simulations \cite{Rajiv0,Rajiv1,Rajiv2} for amorphous materials. \par
In this letter, quantitative results are presented, which show that $\xi _c$ indeed decreases with the crack velocity, and lead to an estimate of $\phi $. Variations 
of the crossover length $\xi_c$ with the average crack velocity are presented here both for the fatigue 
fracture of the Ti$_3$Al-based Super$\alpha_2$ intermetallic alloy and for the stress corrosion fracture of soda-lime silica glass.\par

Two notched compact tension specimens of Super$\alpha _2$ are broken in fatigue.
Fatigue tests are carried out using an electro-servohydraulic testing machine, operating under load control.
The test is performed in air with a constant stress ratio $R=\sigma _{min} / \sigma _{max}=0.1$ ($\sigma _{max}$ and $\sigma_{min}$ are respectively the minimum and the maximum stresses),  at a frequency
$f=30 Hz$. The evolution of the crack length $a$ with time is measured with the potential drop method \cite{potdrop}. The fracture surfaces are
observed for 4 different velocities spanning a wide range, using both an atomic force microscope (AFM) and
a standard scanning electron microscope (SEM). The SEM observations consist in cutting and polishing the NiPd-plated fractured samples perpendicularly
to the direction of crack propagation and registering images of the profiles at various magnifications for each of the 4 regions (see \cite{Turin}). For AFM observations, five profiles
of length 1 $\mu$m are registered in each region.\par

\begin {figure}
\psfig {figure=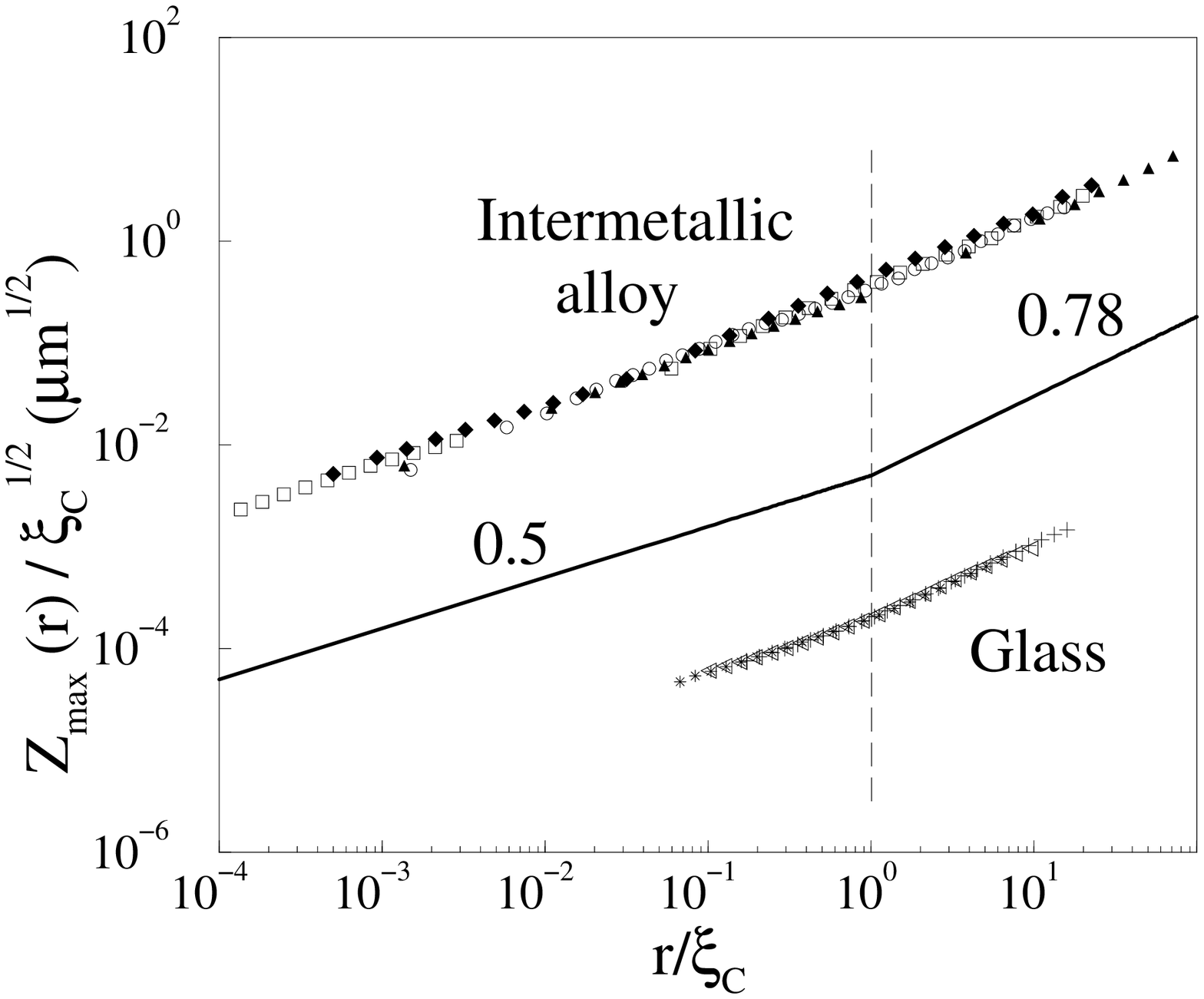,width=8 cm}
\end {figure}
\parbox {8 cm}{{\bf Fig 1:} $Z_{max} (r)/\sqrt{\xi_c} $ is plotted against $r/\xi_c$ for the two materials separately. Note that in these reduced units, the plots corresponding to the various velocities collapse on the same curve. Although the crossover regions are quite different for the two materials, the asymptotic regimes are well described by power laws with exponents 0.5 ($r/\xi_c \ll 1$) and 0.78 ($r/\xi_c \gg 1$). }\vskip 0.2 cm

Fracture surfaces of soda-lime silica glass have been prepared by controlling the crack propagation with a four points bending system. After the initial propagation, which allows to relax all residual stresses, the plate is properly loaded in order to obtain the required average crack velocity. This velocity is measured by imaging the crack tip with AFM at different times. The humidity rate has been measured, and kept between 37 and 41$\%$. The controlled crack propagation is maintained over a distance of about 30$\mu$m, so that fracture surfaces can be easily probed with AFM. Crack velocities range from $2\  10^{-9}$ to $10^{-7}$ ms$^{-1}$. Ten AFM height profiles of length 1.5$\mu$m are registered perpendicularly to the direction of crack propagation, on three samples, and along this direction for four other specimens. As it will be shown in the following, no significant anisotropy could be detected.\par

In order to determine the roughness exponents $\zeta$ and the crossover length $\xi _c$ of the profiles recorded, the
Hurst method is used \cite{SCHM}, where the following quantity is computed: $ Z_{max} (r)=
\langle max[z(r')]_{r_o<r'<r_o+r}   
- min[z(r')]_{r_o<r'<r_o+r} \rangle _{r_o}$, $Z_{max}(r) \propto r^\zeta$. $r$ is the width of the window, and $Z_{max}(r)$ is the difference between the maximum and the minimum heights $z$
within the window, averaged over all possible origins $r_o$ of the window belonging to the profile. \par

\begin {figure}
\centerline {\psfig {figure=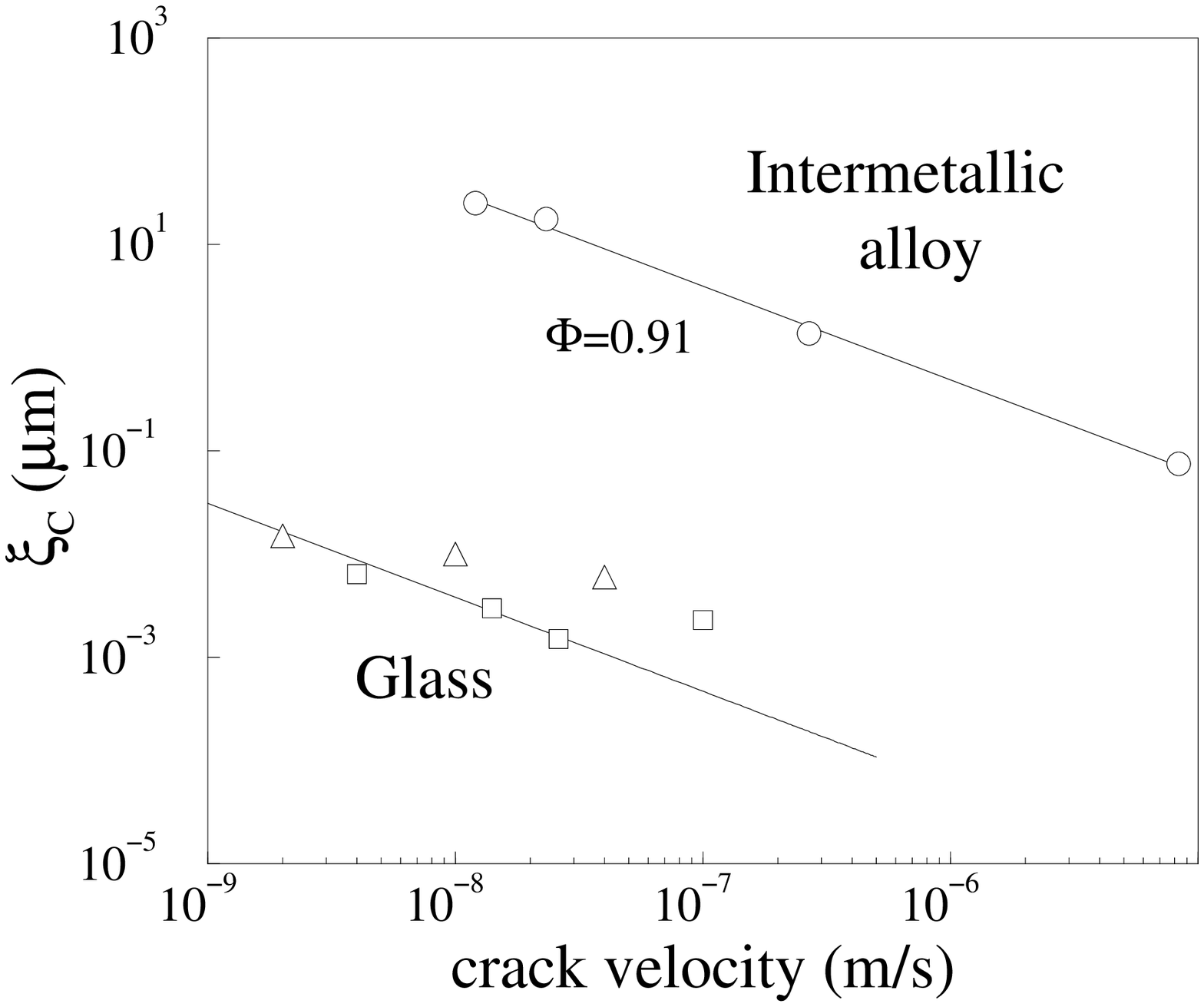,width=8 cm}}
\label {decroit}
\end {figure}
\parbox {8 cm}{{\bf Fig 2:} Evolution of the cross over length $\xi_c$ with the crack velocity for the Super$\alpha _2$ ($\circ $) and
for soda-lime silica glass ($\bigtriangleup$/$\Box $: perpendicular/parallel to the direction of crack propagation). $\xi_c$ is plotted versus $v$ on a log-log plot, exhibitting a power-law dependence with an exponent $\phi \simeq 0.91$.}\vskip 0.2 cm

As far as the Super$\alpha _2$ is concerned, the simultaneous use of SEM and AFM allows for an observation of the fracture
surfaces over 5 or 6 decades of length scales. Within the whole range of observations, $Z_{max} (r)$ is very well fitted by the sum of two power laws, $Z_{max} (r)=A((r/\xi _c)^{0.5}+(r/\xi _c)^{0.78})$. The small and large length scales roughness indices $-$ 0.5 and 0.78 respectively $-$, are chosen to fit with the results of previous experiments \cite{Daguier}. On the contrary, the crossover between the two regimes is much sharper in the case of glass (which seems to be the case for other amorphous materials \cite{Rajiv1,Rajiv2}), and in this case, the crossover length $\xi_c$ is determined as the intersection of the two asymptotic power law regimes with exponents 0.5 and 0.78. Once the crossover lengths have been determined in each case, it is possible to plot $Z_{max}$ as a function of $r/\xi _c$. In Fig. 1, the curves $Z_{max} (r)/\sqrt {\xi_c}$ relative to each material are plotted as a function of $r/\xi_c$ and shown to collapse on the same master curve. In both cases, the asymptotic regimes are well described by power laws with exponents 0.5 at small length scales ($r/\xi_c \ll 1$), and 0.78 at large length scales ($r/\xi_c \gg 1$). In other words, one can write:
$$Z_{max}(r)\propto r^{0.5}\  f({r\over \xi_c}) \eqno (1)$$
with $f(x\rightarrow 0)\sim 1$ and $f(x \gg 1) \sim x^{0.28}$, showing that the amplitude of the small length scales contribution is independent of crack velocity. 
\par

The results obtained on materials as different as an intermetallic alloy and a glass thus confirm previous observations \cite {Daguier,EBSN}, where the short and large length scales regimes were interpreted, respectively, as a ``quasi-static" and a ``dynamic" regime.\par
As it can be seen in Fig. 2, $\xi _c$ decreases with the crack velocity $v$ in both cases, although the measured values of $\xi _c$ are approximately 1000 times larger in the case of the Super$\alpha _2$ than in the case of glass. The experimental results are compatible with a power law decrease $v^{-\phi}$ in both cases. However, the estimated value of $\phi $ is close to unity instead of 3. Note that the values of $\xi_c$ measured for glass for the higher velocities might be overestimated: in this case the precision is very bad, since $\xi_c$ is of the order of some nanometers, i.e. close to the limit resolution of the AFM.\par
\begin {figure}
\centerline {\psfig {figure=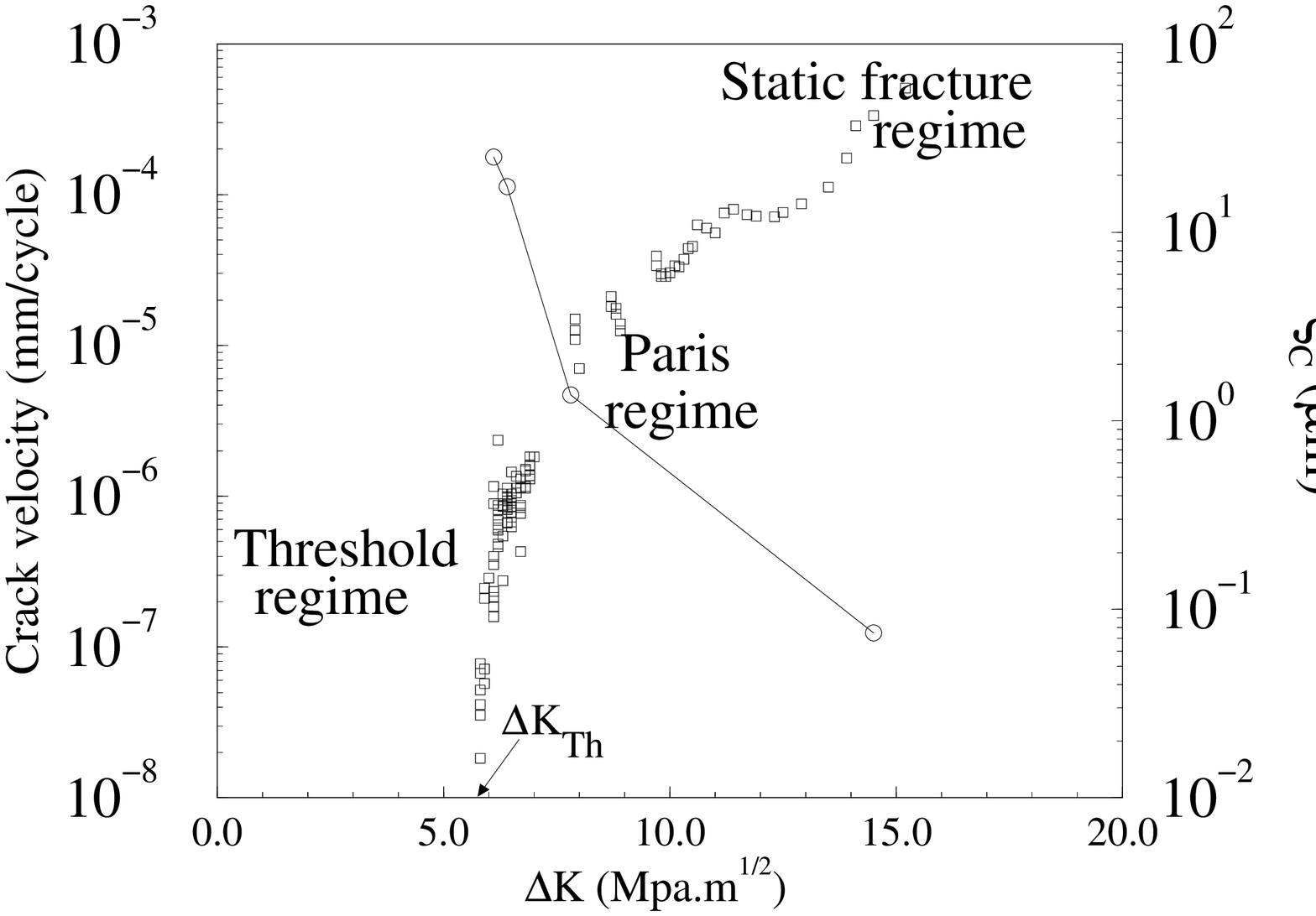,width=8 cm}}
\label {interm}
\end {figure}
\parbox {8 cm}{{\bf Fig 3:} Super$\alpha_2$: the fatigue crack velocity ($\Box$) as well as the cross over 
length $\xi_c$ ($\circ$) are plotted against the stress intensity factor $\Delta K$. The threshold value $\Delta K_{Th}$ 
is indicated.}\vskip 0.2 cm

In Fig. 3, both $\xi_c$ and $v$ are plotted against the stress intensity factor  $\Delta K=(\sigma _{max}- \sigma_{min})\sqrt{a}$. The fatigue crack growth in this regime is widely known as intermittent \cite {lankford,bathias}.
In this regime, the crack tip opens and closes many times before it can
extend over a small distance. This process is repeated several
times and causes incremental crack advance. The number of cycles required to
get the crack to advance decreases as $\Delta K$ increases, and the crack motion is more and more continuous, 
microstructural obstacles being efficient at smaller length scales. At a given time, the force $F$ exerted on 
the fracture front is proportional to $\Delta K$, while the threshold force $F_c$ is proportional to 
$\Delta K_{Th}$ (defined in Fig. 3). The frequency of oscillation of these forces being far more rapid than crack 
propagation, the average force only can be considered, which legitimates the analogy with the above-quoted models. In 
the case of glass, $F$ is proportional to the stress intensity factor $K$, while $F_c$ is proportional to the threshold $K_{Th}$. 
Preliminary results indicate that, in the sub-critical regime, the crack velocity is not uniform, and intermittency is likely 
to occur. Thus, in both cases, the pinning/depinning scenario is qualitatively satisfactory.\par

Fig. 4 shows the evolution of the crack velocity $v$ as a function of $\Delta K- \Delta K_{Th}$ for the 
Super$\alpha _2$ and as a function of $K-K_{Th}$ for glass. In the case of the Super$\alpha _2$, experimental measurements 
reveal a power law increase without any change between the so-called ``threshold" and Paris regimes. On the contrary, when 
static fracture occurs, a clear deviation from the power law can be observed for high values of $\Delta K- \Delta K_{Th}$. A 
fit of these data gives $\beta \simeq 2$. This value is compatible with the measurements on glass (Fig. 4).\par

\begin {figure}
\centerline {\psfig {figure=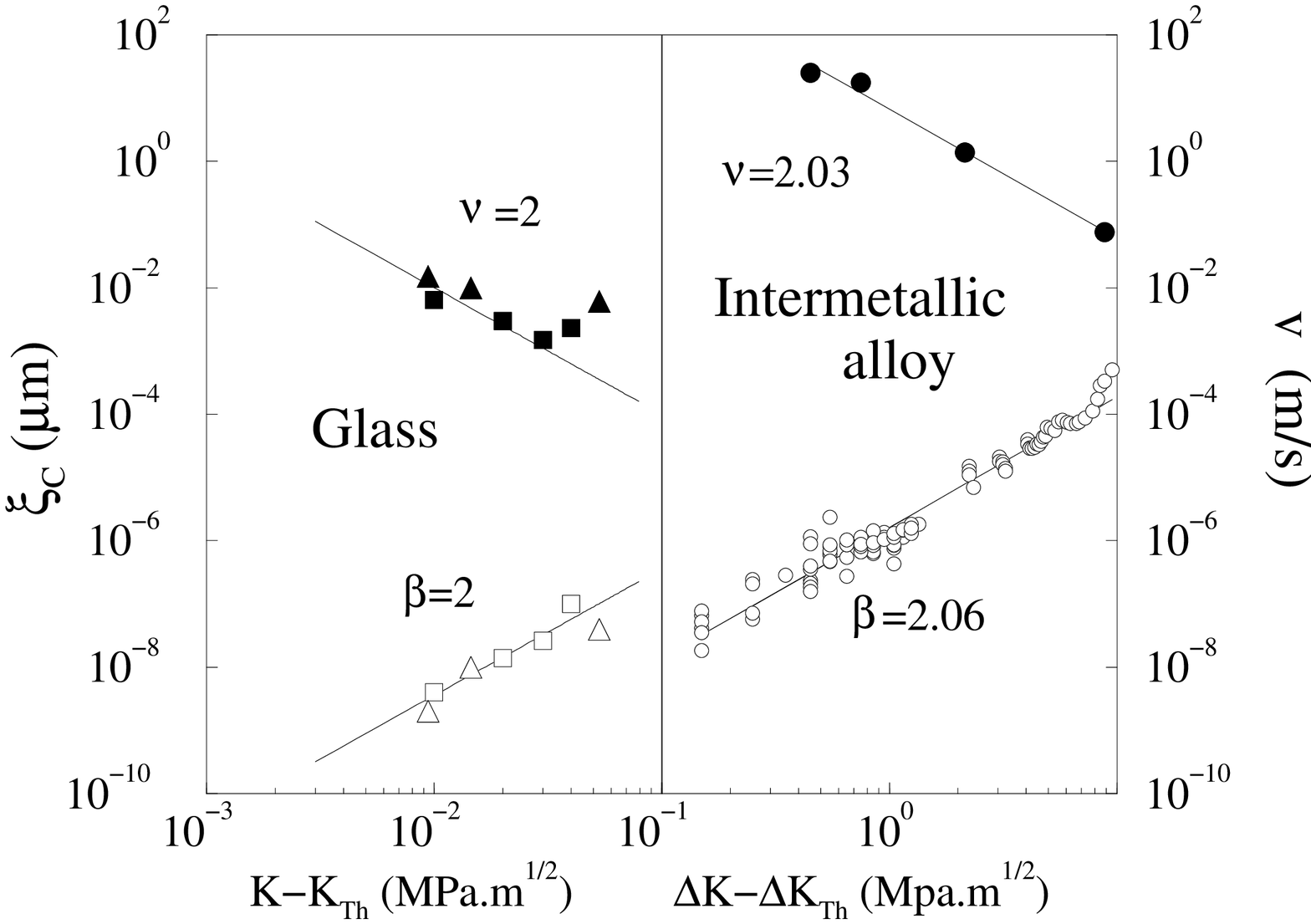,width=8 cm}}
\label {interm}
\end {figure}
\parbox {8 cm}{{\bf Fig 4:} Super$\alpha _2$ ($\circ$): the fatigue crack velocity (white symbols) is plotted 
versus $\Delta K-\Delta K_{Th}$ on a log-log plot, as well as the crossover length $\xi_c$ (black symbols). \par
Glass ($\bigtriangleup$/$\Box $  perpendicular/parallel to the direction of crack propagation): the crack velocity 
is plotted as a function of $K-K_{Th}$ (white symbols), as well as $\xi_c$ (black symbols).}\vskip 0.2 cm

In Fig. 4, $\xi_c$ is plotted also for both materials. A power law decrease can be observed, and the 
fit of the data relative to the metal gives $\nu \simeq 2$, compatible with the results on glass.\par
One can note that $\phi=\nu /\beta \simeq 1$. On the other hand, it is expected that the 
exponent $n$ characterising the range of interactions ($n=2$ for short range forces \cite{Fisher} 
and $n=1$ for long range ones \cite{ertaskardar3}) is related to the exponents $\nu$ and $\zeta_{\parallel}$ through the relation:
$$n=\zeta_{\parallel}+{1\over \nu} \eqno (2)$$
where $\zeta_{\parallel}$ is the in-plane roughness exponent \cite{inplane}. $\zeta_{\parallel}$ was determined previously on the Super$\alpha _2$ \cite{inplane}, and shown to be close to $\zeta_{\parallel}\simeq 0.54$. This leads to a value of $n\simeq 1.03$ very close to unity, as expected for elastic interactions \cite{ertaskardar3,SNL,Maya}. Note that the 2d model of Thomas and Paczuski\cite{Maya} leads to $\zeta=0.5$ and $\nu=2$, but also to $\beta=1$.\par
Knowing $\beta $ and $\nu $, one can in principle deduce the value of the dynamic exponents $z_{\parallel}$ and $z_{\perp}$ describing the short time evolution of the front, repectively in the direction of crack propagation and perpendicularly to it. $z_{\parallel}=\zeta_{\parallel}+{\beta \over \nu}$ leads to $z_{\parallel} \simeq 1.5$, while $z_{\perp}=z_{\parallel}+{1\over \nu}$
should indicate that $z_{\perp}\simeq 2$. Hence, perturbations on the crack front are diffusive perpendicularly to the direction of crack propagation, while they are slightly hyper-diffusive along this direction.\par
For ductile materials as the Super$\alpha _2$, the plastic zone size $R_{plast}$ should be a relevant length scale as well. Although the same behaviour is observed within the whole range of $\Delta K$s, it can be noted that $R_{plast}$ overpasses $\xi_c$ for the two experiments corresponding to higher velocities.\par

The transition between the threshold regime and the Paris regime might be associated respectively with $\xi_c > R_{plast}$ and
$\xi_c < R_{plast}$. It has been shown in Fig. 2 that a coarser microstructure gives rise to a higher $\xi_c$ for the same crack velocity. On the other hand, Yoder {\sl et al} \cite {yoder}
showed that, in the case of a titanium-based alloy, a coarser microstructure corresponds to a higher value of $\Delta K$ 
at which the transition occurs. This result supports the idea
that the transition between the threshold regime and the Paris regime could be associated to a competition between $R_{plast}$
and $\xi_c$. Note also that in the case of glass, the plastic zone size has been estimated \cite{EG} to be of the order of some nanometers, i.e. of the order of magnitude of $\xi_c$.\par
Further experiments on different materials are needed in order to confirm the general character of the pinning/depinning scenario, and to allow for a more precise determination of the critical exponents. In order to investigate the role of plasticity, experiments will be performed on an aluminium alloy, for which plastic zones are much larger than in the case of the Super$\alpha _2$.\par

{\bf Acknowledgements}: The authors are particularly indebted to S. Nav\'eos and G. Marcon for their technical help, 
and to J.-P. Bouchaud, E. Orignac and S. Roux for enlightening discussions. \\

\end {multicols}
\end {document}